\documentclass[apj,twocolumn]{emulateapj}
\usepackage{amsmath}
\usepackage{txfonts}
\usepackage{amssymb}
\usepackage{graphicx}
\usepackage{textcomp}
\usepackage{color}
\usepackage{enumerate}
\newcommand{\dsolar}{$\rm d_{\odot}$}
\renewcommand\email\texttt

\begin{document}
\shorttitle{Extra-tidal features around Whiting 1}
\shortauthors{Nie et al.}
\title{Searching extra-tidal features around the Globular Cluster Whiting 1}

\author{
Jundan Nie\altaffilmark{1,$\star$},
Hao Tian\altaffilmark{2,$\star$},
Jing Li\altaffilmark{3},
Chao Liu\altaffilmark{2},
Martin C. Smith\altaffilmark{4},
Baitian Tang\altaffilmark{5},
Julio A. Carballo-Bello\altaffilmark{6},
Jun Ma\altaffilmark{1,7},
Haijun Tian\altaffilmark{8},
Jiaxin Wang\altaffilmark{9},
Zhenyu Wu\altaffilmark{1},
Xiyan Peng\altaffilmark{4},
Jiali Wang\altaffilmark{1},
Tianmeng Zhang\altaffilmark{1},
Xu Zhou\altaffilmark{1},
Zhimin Zhou\altaffilmark{1},
Hu Zou\altaffilmark{1}
}
\altaffiltext{1}{CAS Key Laboratory of Optical Astronomy, National Astronomical Observatories, 
Chinese Academy of Sciences, 
Beijing 100101, PR China; \email{jdnie@nao.cas.cn;tianhao@nao.cas.cn}}
\altaffiltext{2}{Key Laboratory of Space Astronomy and Technology, 
National Astronomical Observatories, 
Chinese Academy of Sciences, 
Beijing 100101, PR China;}
\altaffiltext{3}{hysics and Space Science College, China West Normal University, 1 ShiDa Road, Nanchong 637002, China;}
\altaffiltext{4}{Shanghai Astronomical Observatory, Chinese Academy of Sciences, Nandan Road, Shanghai 200030, China;}
\altaffiltext{5}{School of Physics and Astronomy, Sun Yat-sen University, Zhuhai 519082, People's Republic of China;}
\altaffiltext{6}{Instituto de Alta Investigaci\'on, Universidad de Tarapac\'a, Casilla 7D, Arica, Chile;}
\altaffiltext{7}{College of Astronomy and Space Sciences, University of Chinese Academy of Sciences, Beijing 100049, China;}
\altaffiltext{8}{Center for Astronomy and Space Sciences, China Three Gorges University, Yichang 443002, China;}
\altaffiltext{9}{School of Science, Chongqing University of Posts and Telecommunications, Chongqing 400065,China;}
\altaffiltext{$\star$}{Email: \email{jdnie@nao.cas.cn;tianhao@nao.cas.cn};}

\begin{abstract}
Whiting 1 is a faint and young globular cluster in the halo of the Milky Way, and was 
suggested to have originated in the Sagittarius spherical dwarf galaxy (Sgr dSph). In this paper,
we use the deep DESI Legacy Imaging Surveys to explore tentative spatial connection between Whiting 1 and the 
Sgr dSph. We redetermine the fundamental parameters of Whiting 1 and use the best-fitting isochrone 
(age $\tau$=6.5 Gyr, metalicity Z=0.005 and \dsolar=26.9 kpc) to construct a theoretical matched
filter for the extra-tidal features searching. Without any smooth technique to the matched filter
density map, we detect a round-shape feature with possible leading and trailing tails on either side
of the cluster. This raw image is not totally new compared to old discoveries, but confirms that no more large-scale features can be
detected under a depth of $r<=$22.5 mag. In our results, the whole feature stretches 0.1-0.2 degree along the orbit of Whiting 1, which gives a much larger area than
the cluster core. The tails on both sides of the cluster align along the orbital direction of the Sgr dSph as well as
the cluster itself, which implies that these debris are probably stripped remnants of Whiting 1 by the Milky Way. 
\end{abstract}

\section{Introduction}\label{introduction}
Under the hierarchical formation paradigm, the halo, at least partly, was formed
through the continuous merging of satellites, such as clusters and dwarf galaxies 
\citep{1996ApJ...465..278J}. The remnants of those events are left in the form of
hundreds of dispersed tidal debris or stellar streams
\citep{1999MNRAS.307..495H,1999ASPC..165...89H,2003MNRAS.339..834H}, such as debris 
for the Gaia-Enceladus-Sausage
\citep{2018MNRAS.478..611B,2018ApJ...863..113H,2018Natur.563...85H,2018ApJ...856L..26M},
for the Sequoia galaxy \citep{2019MNRAS.488.1235M}, and the Helmi streams
\citep{1999Natur.402...53H}, the Sagittarius streams
\citep{1994Natur.370..194I,2010ApJ...718.1128L}, the tidal streams of Elqui, Aliqa Uma,
and Chenab, etc. discovered by the Dark Energy Survey \citep{2018ApJ...862..114S}. 
Such phenomena are found not only in the Milky Way but also in the M31, e.g.,
a vast thin plane of satellites was discovered by \citet{2013Natur.493...62I} in 
this galaxy. Those merging events are important evidence to constrain Milky Way
formation model. They also provide tracers to estimate Milky Way mass distribution
\citep{2001ApJ...551..294I,2004MNRAS.351..643H,2005ApJ...619..807L,2010ApJ...714..229L,
2015ApJ...803...80K}.

The Sagittarius stream is a perfect example to study the merging of a dwarf galaxy with
the Milky Way. First, the whole stream has orbited several times the Galaxy,  which
clearly indicates the path of the Sagittarius dwarf spheroidal galaxy (Sgr dSph) around
the Milky Way. \citet{2014MNRAS.437..116B} studied the potential of the Milky Way using
its precession and found that the two tails of the Sgr evolve in different ways. Second,
the Galactocentric distances of the member stars vary from $\sim$10 to 120 kpc,
allowing to explore the outer layers of our Galaxy. Since \citet{2003ApJ...599.1082M}
revealed the full-sky distribution of the leading and trailing arms of the Sgr stream,
more and more substructures in detail have been discovered, e.g., \citet{2017ApJ...844L...4S} 
traced the Sgr stream with RR Lyrae stars and found the Sgr Spur structure with Galactocentric 
distance larger than 100 kpc. \citet{2019ApJ...874..138L} also found the same structure with 
M-giant stars selected from LAMOST \citep{2012RAA....12.1197C,2012RAA....12..723Z}.
\citet{2018arXiv180901102B} used RR Lyrae stars to trace the Sgr stream and detected
three additional substructures, with the nearest one located at \dsolar=10.4 kpc,
coincides with the simulation model of \citet{2010ApJ...714..229L}
and the findings of \citet{2010ApJ...721..329C}. Third, the Sgr stream is not only the
debris of the accretion by the Milky Way but also the host of group mergers, which 
contributes to the building up of the Galactic globular cluster system, i.e., a fraction
of globular clusters have been found to be certainly or possibly originally formed in
the Sgr dSph and scattered along the stream arms, such as Arp 2, M 54, NGC 5634, Terzan
8, Berkeley 29, Pal 12, Terzan 7, and Whiting 1
\citep{2000AJ....120.1892D,2002ApJ...573L..19M,2003AJ....125..188B, 2008AJ....136.1147B,
2010ApJ...718.1128L,2014MNRAS.445.2971C,2019A&A...630L...4M,2019MNRAS.484.2832V}.

As one of the intriguing globular clusters associated with the Sgr stream, 
Whiting 1 (RA=02:02:57, Dec=-03:15:10) is a faint and young cluster in the 
halo of the Milky Way. It was firstly discovered by \citet{2002ApJS..141..123W},  
and was suggested as a well-resolved open cluster with a size of
$1.2\arcmin\times1.0\arcmin$. Later, \citet{2005ApJ...621L..61C} used new photometric
data to estimate its fundamental parameters. For the first time, they gave an age
$\tau\sim$5 Gyr, metallicity $Z\sim$0.001([Fe/H]=-1.20), heliocentric distance
\dsolar$\sim$45 kpc, and classified this cluster as an unusually young halo globular
cluster.  With the help of deep Very Large Telescope (VLT) photometry, \citet{2007A&A...466..181C} estimated 
an age of $\tau$=6.5 Gyr, a metallicity $Z$=0.004 ([Fe/H]=-0.65), and a distance of
\dsolar=29.4 kpc for Whiting 1. Meanwhile, using high-resolution spectra from the
Magellan telescope, they obtained the first estimate of its radial velocity, set at
-130.6 km s$^{-1}$. Besides optical photometry, near-infrared photometric data 
(relatively shallower) were also obtained by \citet{2015MNRAS.446..730V}, and they 
gave slightly different cluster parameters, i.e., age of 5.7 Gyr, metallicity of
0.006([Fe/H]=-0.5), and distance of 31.3 kpc. Based on their measurements,  the authors
claimed that Whiting 1 was a young and moderately metal-rich globular cluster.
 
Due to the spatial location (embedded in the Sgr trailing stream) and its exceptional
age and metallicity, the formation and evolution of Whiting 1 has drawn much attention.
\citet{2007A&A...466..181C} used their photometric and spectroscopic data to assess 
the association between Whiting 1 and the Sgr dSph. They find that the distances and
radial velocities of the two systems are comparable. Besides, the age-metallicity
relation of Whiting 1 is consistent with that of the Sgr dSph. All these imply that
Whiting 1 is probably originated within the Sgr dSph. Similar results were given by
\citet{2010ApJ...718.1128L}, who compared the kinematic properties of globular clusters
with the dynamics of the Sgr dSph, and associated Whiting 1 with the trailing stream of
the Sgr dSph. \citet{2017MNRAS.467L..91C} used the 8.2m VLT to obtain
spectra of Sgr stream candidate stars around Whiting 1 and obtained a radial velocity
component of -130 km s$^{-1}$, which indicates that the Sgr stream stars have similar
kinematic properties with Whiting 1. All above compatible properties between the two
systems seem to form an evolution path for Whiting 1: this young cluster is likely to
have originally formed in the Sgr dSph and  gradually grew into the current shape by 
the accretion of the Milky Way, and now behaving similar properties with its
progenitor. 

If Whiting 1 is indeed associated with the Sgr dSph, its morphology should show some
tentative spatial connection with its progenitor. From the King-profile fit of the surface 
density profile of Whiting 1 \citep{2007A&A...466..181C}, this cluster seems to have 
extra members outside its tidal radius. By using the Matched Filter (MF) analysis of the 
stars around Whiting 1, \citet{2017MNRAS.467L..91C} found that the structure of Whiting 1 
is elongated in the opposite direction, which likely aligns with the orbit of the Sgr dSph. 
This agrees with the picture that Whiting 1 was recently accreted by the Milky Way as 
part of the Sgr dSph. However, in their density map, only the nearby tail-like feature 
with a size $<0.1^{\circ}$ was discovered. In another study by \citet{2018MNRAS.476.4814S}, 
which mainly focused on finding dwarf galaxy remnants around globular clusters, they
presented the density map of Whiting 1 with the $k$-nearest neighbor density estimator.
However, in their analysis, the elongation suggested by \citet{2017MNRAS.467L..91C} is
not observed. To verify which picture is true, as well as to better understand
the origin of this cluster, it is quite necessary to search for extra-tidal features around 
Whiting 1.  If Whiting 1 is really disrupted by the Galaxy potential,
the stripped stars would form tidal tails surrounding the cluster, and these remnants
should be observed on the opposite side of the cluster, and probably, with deep data
these remnants could be observed in a larger-structure scale. 

In this paper, we use a photometric sky survey, DECam Legacy Survey (DECaLS), which is
one of the Legacy Surveys of the Dark Energy Spectroscopic
Instrument\footnote{https://www.desi.lbl.gov} (DESI) imaging, to search for extra-tidal
features around Whiting 1 and look for their morphological correlation with the Sgr
dSph. In Section \ref{data_method}, we describe the survey data and the method. We
analyze the results in Section \ref{result} and make the conclusion in the Section \ref{summary}.

\section{Data and Method}\label{data_method}
\subsection{DECaLS data}
We use DECaLS DR9 data to search for extra-tidal features
around the globular cluster Whiting 1. DECaLS\footnote{http://legacysurvey.org}
\citep{2019AJ....157..168D} is one of the imaging Legacy Surveys of the DESI. 
It utilizes the DECam imager of the 4 m Blanco telescope at the Cerro Tololo
Inter-American Observatory, National Optical Astronomy Observatory (NOAO), to 
survey about 9000 square degrees of the equatorial sky 
(-50$^{\circ}$$<$RA$<$270$^{\circ}$ and -18$^{\circ}$$<$Dec$<$30$^{\circ}$) with 
$g$ and $r$ filters. The DECam camera contains 62 CCDs with 520 megapixels and 
images 3 square degrees at a 0.263$\arcsec$/pixel resolution. The global 5$\sigma$ depth 
are 24.0 and 23.4 mag for $g$ and $r$ bands, respectively. For individual fields,
the depth might vary due to observing condition.

DECaLS DR9 provides five morphological magnitude types, and the point-spread function (\textit{PSF}) 
magnitude is provided only for objects best fit as point sources. We are only 
interested in stellar sources, so all \textit{PSF}
magnitudes with 16$<g<$24, and $g,r$ uncertainties $<$0.2 mag (equivalent to signal-to-noise
ratio or S/N
$>5\sigma$) are extracted from the catalog.We constrain the working area to a field of
$10^{\circ}\times10^{\circ}$ centered in Whiting 1. Note that this searching area 
is much larger than the working fields of previous studies (e.g.,
$1.0^{\circ}\times1.0^{\circ}$ in \citet{2017MNRAS.467L..91C} and
\citet{2018MNRAS.476.4814S}). All the magnitudes are extinction corrected by using
E(B-V) values from the dust maps of \citet{1998ApJ...500..525S} with extinction
coefficients of 3.214 and 2.165 for $g$ and $r$ bands \citep{2019AJ....157..168D},
respectively. In the text below, if not specified, we retain to use $g,r$ for the
extinction-corrected magnitudes.

\begin{figure}
\center
\includegraphics[width=0.5\textwidth]{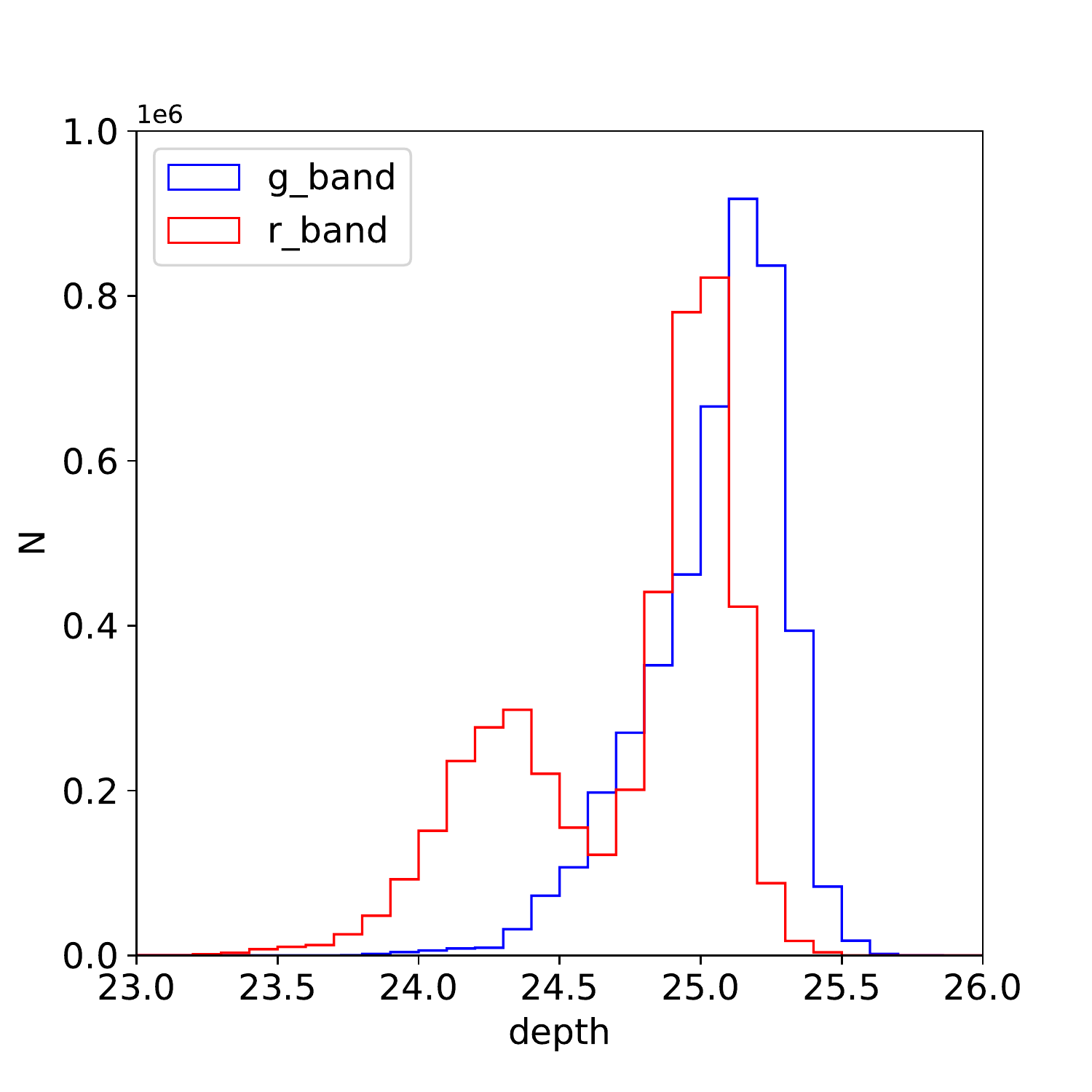}
\center
\caption{The depth distribution for all point sources in our working data. 
Blue histogram is for $g$ band, and red histogram is for $r$ band.}
\label{fig_depth}
\end{figure}

To assess the spatial fluctuations due to the survey completeness, we
analyze the depth of DECaLS data around Whiting 1. In Figure \ref{fig_depth}, we show the
depth distribution for all stars in our working data. As shown in the figure, 
the median depth is much deeper than $g=$24.0 or $r$=23.4 mag.
The depth for all the individual stars are deeper than 23.8 mag in $g$ band and 23.2 mag in $r$ band.
This represents that the 100\% completeness for our working field reaches at 23.8 mag and
23.2 mag for $g$ and $r$ band, respectively. \citet{2021AJ....161...12X} calculates
the completeness of the DECaLS data by the star detection rate, and assesses that the 100\%
completeness reaches at $r=$22.5 mag. To avoid spatial fluctuations due to the survey
completeness, in our extra-tidal feature-detection work, we will use $r=$22.5 mag as 
a magnitude limit.

\subsection{Determination of the fundamental parameters of Whiting 1 }\label{isochrone}
The age $\tau$, metallicity $Z$ and distance modulus $dm$ are the fundamental
parameters for a cluster. The obtaining of these parameters is vital for the
performance of the MF method and membership confirmation.

\begin{figure}
\center
\includegraphics[width=0.5\textwidth]{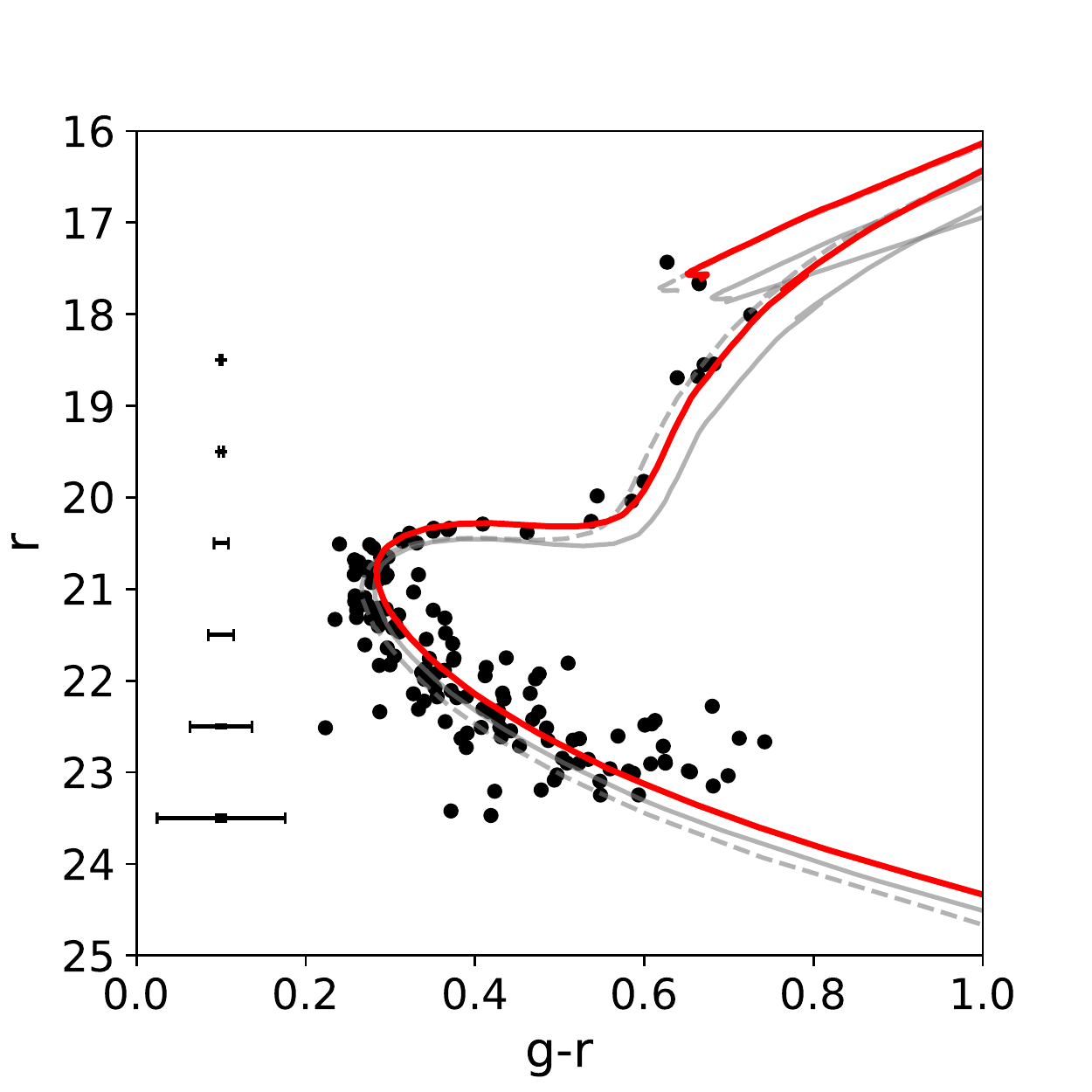}
\center
	\caption{The observational CMD of Whiting 1, ($g-r$) vs. $r$. Black dots denote 
	the extinction-corrected CMD with stars inside the tidal radius (r$_{t}<$1$\arcmin$) of Whiting 1. 
	Typical observational error bars are shown at six magnitude levels. Red line
	denotes our best-CMD fitting with age $\tau$=6.5 Gyr, $z$=0.005, and \dsolar=26.9 kpc, 
	dashed and solid gray lines denote the model fittings from
	\citet{2007A&A...466..181C} and \citet{2015MNRAS.446..730V}, respectively.}
\label{fig_isochrone}
\end{figure}

In this section, we use the DECaLS DR9, which is $\sim$3 magnitude deeper than
the main-sequence turn-off of Whiting 1, to redetermine
the fundamental parameters of Whiting 1. The estimation of the parameter-solution is 
performed by comparing the observational color-magnitude diagram (CMD) to the
theoretical isochrones. A purer CMD,  $(g-r)$ versus $r$, is constructed with stars
within the tidal radius of Whiting 1 ($r_{t}<$1.0$\arcmin$, 
\citep{2005ApJ...621L..61C,2007A&A...466..181C,2015MNRAS.446..730V,2017MNRAS.467L..91C}),
as shown in Figure \ref{fig_isochrone}. A larger tidal radius of $r_{t}=8.4\arcmin$ given by \citet{2018ApJ...860...66M} is not used because their Canada France Hawaii
Telescope (CFHT) observations are deeper than the DECaLS survey. From the DECaLS data, we do not find any blue horizontal-branch stars
in this cluster, which confirms that the cluster is young and moderately metal-rich. 
The theoretical isochrones are obtained from the PARSEC v1.2+COLIBRI S35 evolutionary
tracks \citep{2012MNRAS.427..127B,2014MNRAS.444.2525C,2014MNRAS.445.4287T,
2015MNRAS.452.1068C,2017ApJ...835...77M,2019MNRAS.485.5666P} with magnitudes in the 
DECam bands. No circumstellar dust is applied, and a constant extinction of E(B-V)=0.023
\citep{1998ApJ...500..525S} is used. We keep the extinction constant because only a 
small variation in reddening is observed over the whole Whiting 1 area. The initial mass
function is expected to follow the \citet{1955ApJ...121..161S}  power law. We vary the 
age $\tau$ from 5.5 to 7.5 Gyr with a step of 0.01 Gyr, $Z$ from 0.001 to 0.01 with a
step of 0.001, and $dm$ from 17 to 18 with a step of 0.01 mag. 

We calculate the $\chi^{2}$ for each ($\tau,Z,dm$) set, and the best-fitting model is the
one having the minimal $\chi^{2}$,  with $\tau$=6.5$\pm$0.1 Gyr, $Z$=0.005$\pm$0.001
([Fe/H]=-0.55$\pm$0.09), and $dm$=17.22$\pm$0.01 mag (equivalent to \dsolar=26.9$\pm$0.1
kpc). The best-fit isochrone for the observational CMD is represented by the red solid
line in Figure \ref{fig_isochrone}. To double check, we select the stars within
0.5$\arcmin$ which is half of the tidal radius, and repeat the steps above. A similar
solution is obtained. 

Comparing to previous estimations, our solution is slightly different from
\citet{2015MNRAS.446..730V}, but within 2$\sigma$ deviation of the parameters in
\citet{2007A&A...466..181C}(see their solutions in our reference and fittings in Figure
\ref{fig_isochrone}). Since our parameter determination was done by covering the main
sequence with deeper data, it gives us a more accurate and less-biased parameter
estimation. 

\subsection{The orbit of Whiting 1}\label{orbit}
\begin{figure*}
\centering
\includegraphics[width=0.6\textwidth]{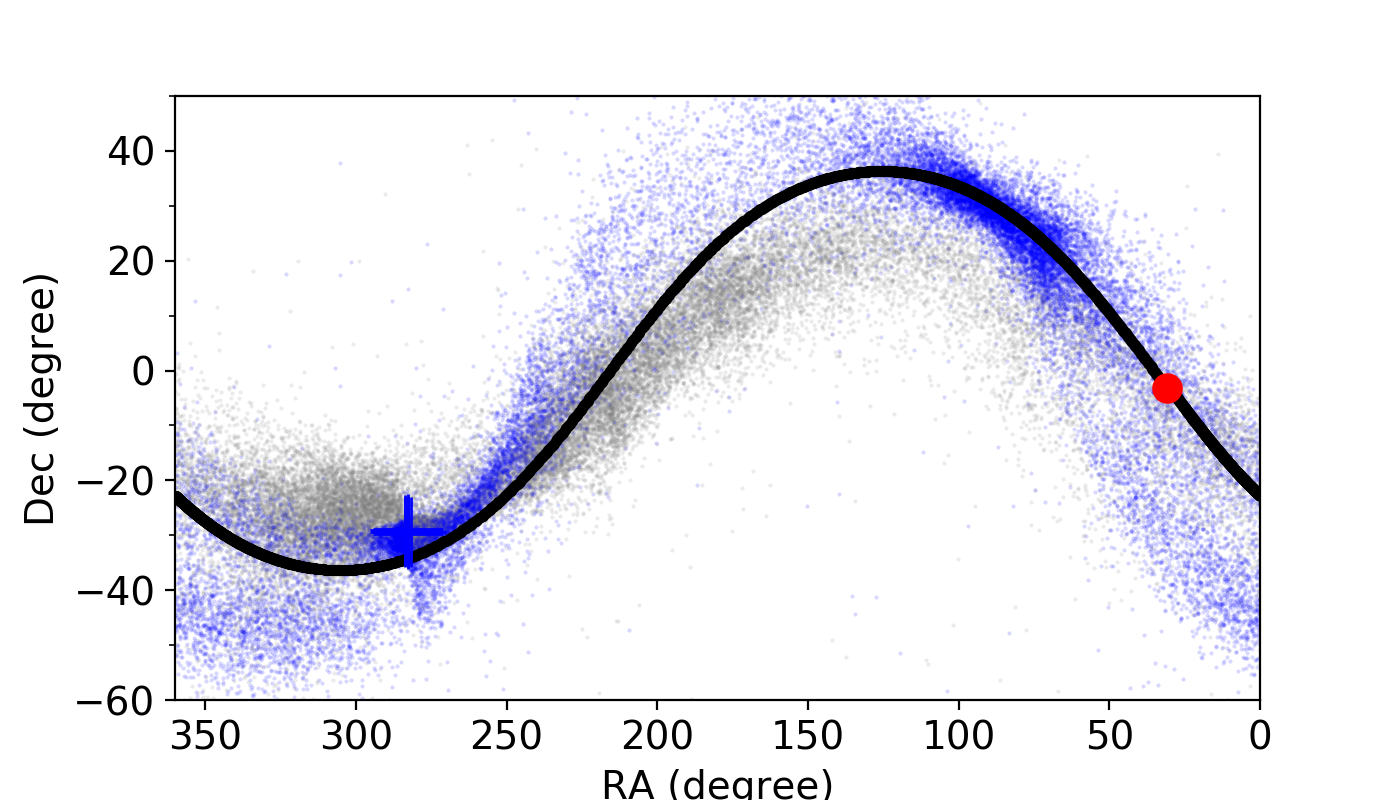}
\includegraphics[width=0.6\textwidth]{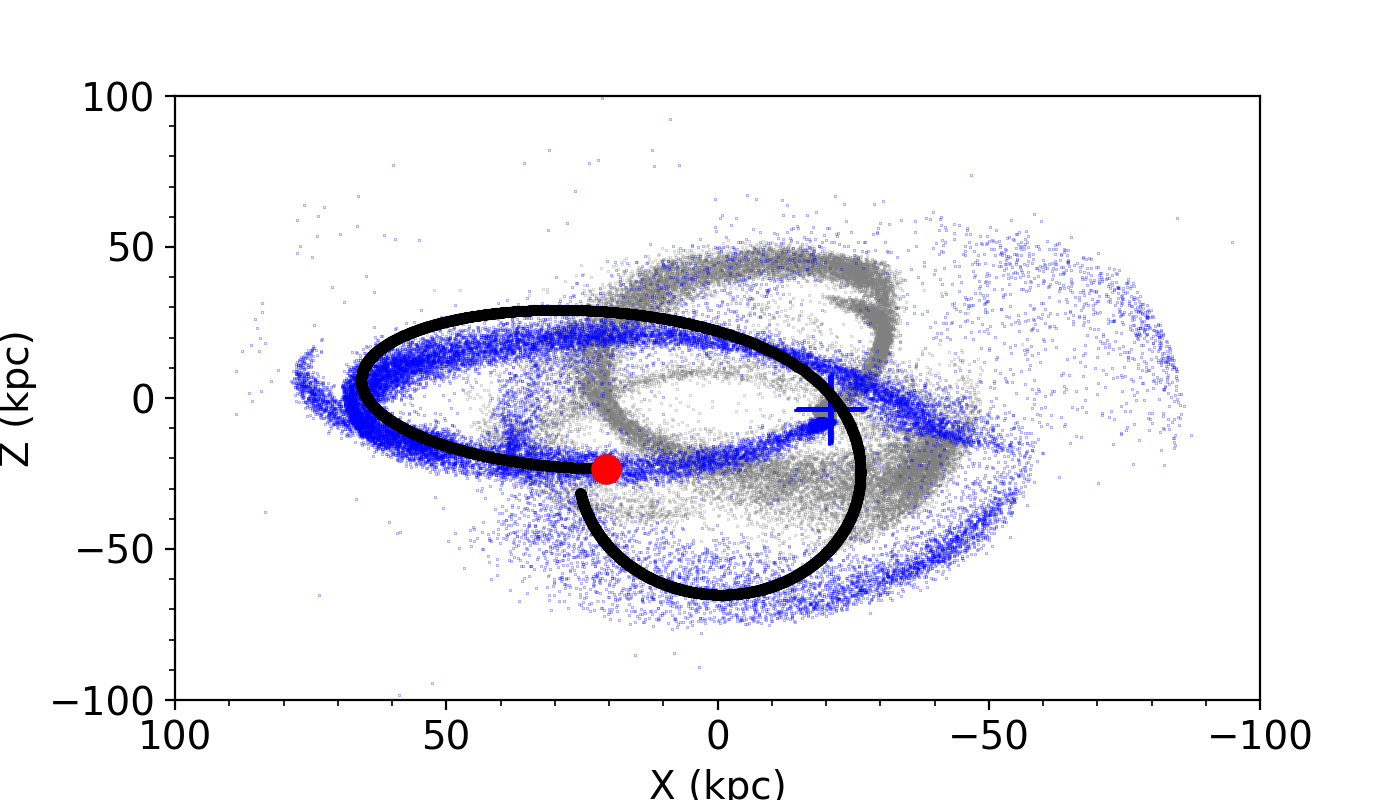}
\includegraphics[width=0.6\textwidth]{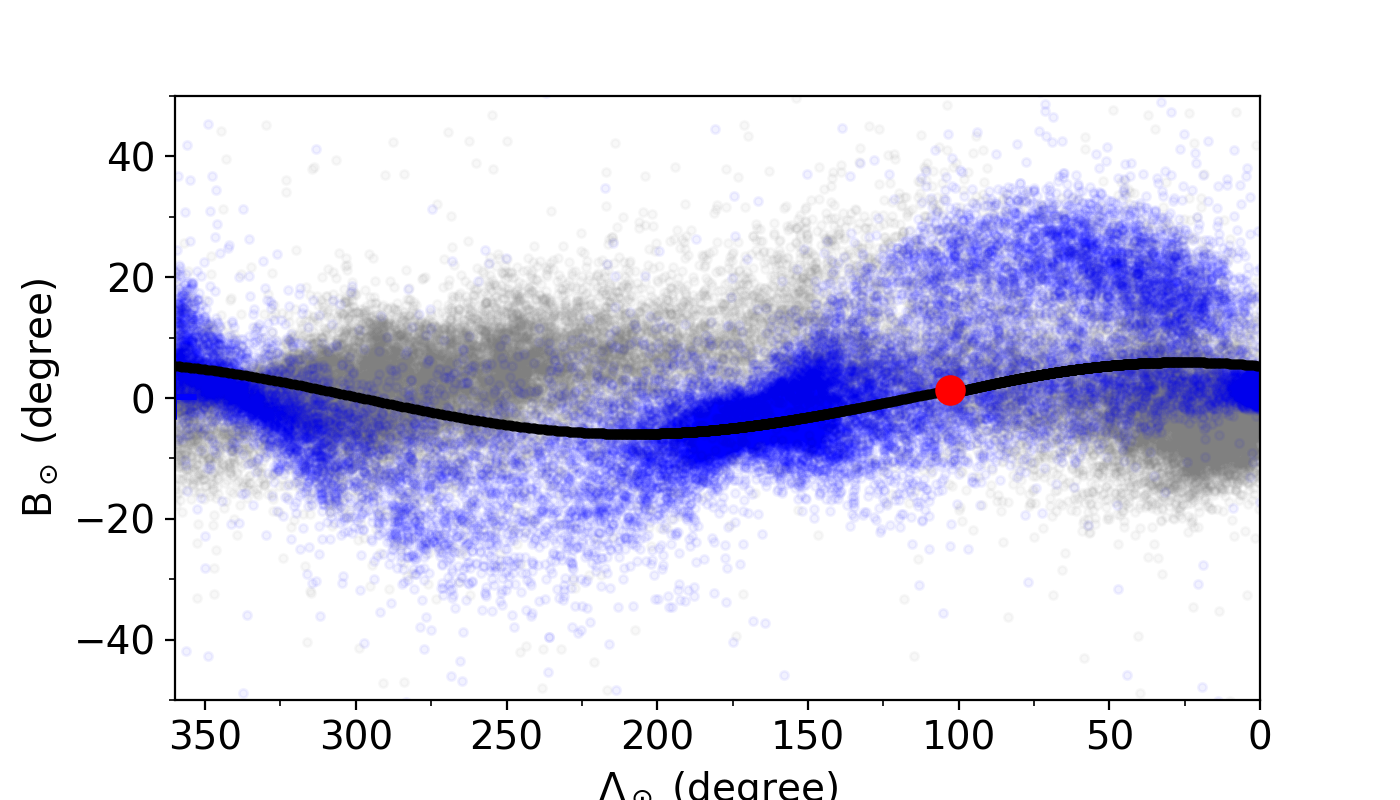}
\caption{The orbit of Whiting 1. The top panel shows the orbit in right ascension and 
declination, the middle panel in Galactic XZ plane, while the bottom panel in a 
coordinate system ($\Lambda_{\odot}$,B$_{\odot}$) aligned with 
the orbit of Sagittarius, as defined in \citet{2003ApJ...599.1082M}. In each panel, 
red dot denotes the current position of Whiting 1, a solid black line denotes the 
integrated orbit of Whiting 1 from present back to 1.95 Gyr ago, a blue plus denotes the 
position of the Sgr dSph, blue and gray dots denote the modeling particles of the Sgr 
trailing and leading streams from the data of \citet{2010ApJ...714..229L}. } 
\label{fig_orbit}
\end{figure*}

With help of Gaia EDR3 \citep{2016A&A...595A...1G,2021A&A...649A...1G}, we are able to 
compute the orbit of the Whiting 1 cluster. The mean proper motion of the stars within 
the tidal radius of the cluster is ($\mu_{\alpha}$,$\mu_{\delta}$)=(-0.1,-2.3) mas/year. 
Together with the sky position, the distance to the Sun, and the radial velocity of 
-130.6 km$^{-1}$, we use the function \emph{orbit} in
\emph{Galpy}\footnote{http://github.com/jobovy/galpy}\normalfont{}\citep{2015ApJS..216...29B} 
to trace back the orbit assuming the \emph{MWPotential2014} potential 
\normalfont{}\citep{2015ApJS..216...29B}. We integrate the orbit back from present to 
-1.95 Gyr, where 1.95 Gyr is about the simulated orbital period of Whiting 1. The orbit of
Whiting 1 is represented by the black solid line in Figure \ref{fig_orbit}, in forms of
three coordinate systems. The dots are the simulated Sagittarius Stream stars from
\citet{2010ApJ...714..229L}, with gray for the leading part and blue for trailing part,
respectively. It clearly shows that the Whiting 1 orbit is quite consistent with that of
the Sagittarius trailing stream in three different coordinate systems. This is not a surprise
because, at the sky location ($\alpha$,$\delta$,\dsolar) of Whiting 1, the observed 3D 
velocity ($\mu_{\alpha}$,$\mu_{\delta}$,RV) of Whiting 1 is very similar to the observations
of the Sgr trailing stream 
\citep{2020A&A...635L...3A,2020A&A...636A.107B,2007A&A...466..181C,2017MNRAS.467L..91C}. 
This consistence provides an extra support that Whiting 1 is associated with the Sgr dSph.

\subsection{Matched filter method}\label{MF}
\begin{figure}
\center
\includegraphics[width=0.5\textwidth]{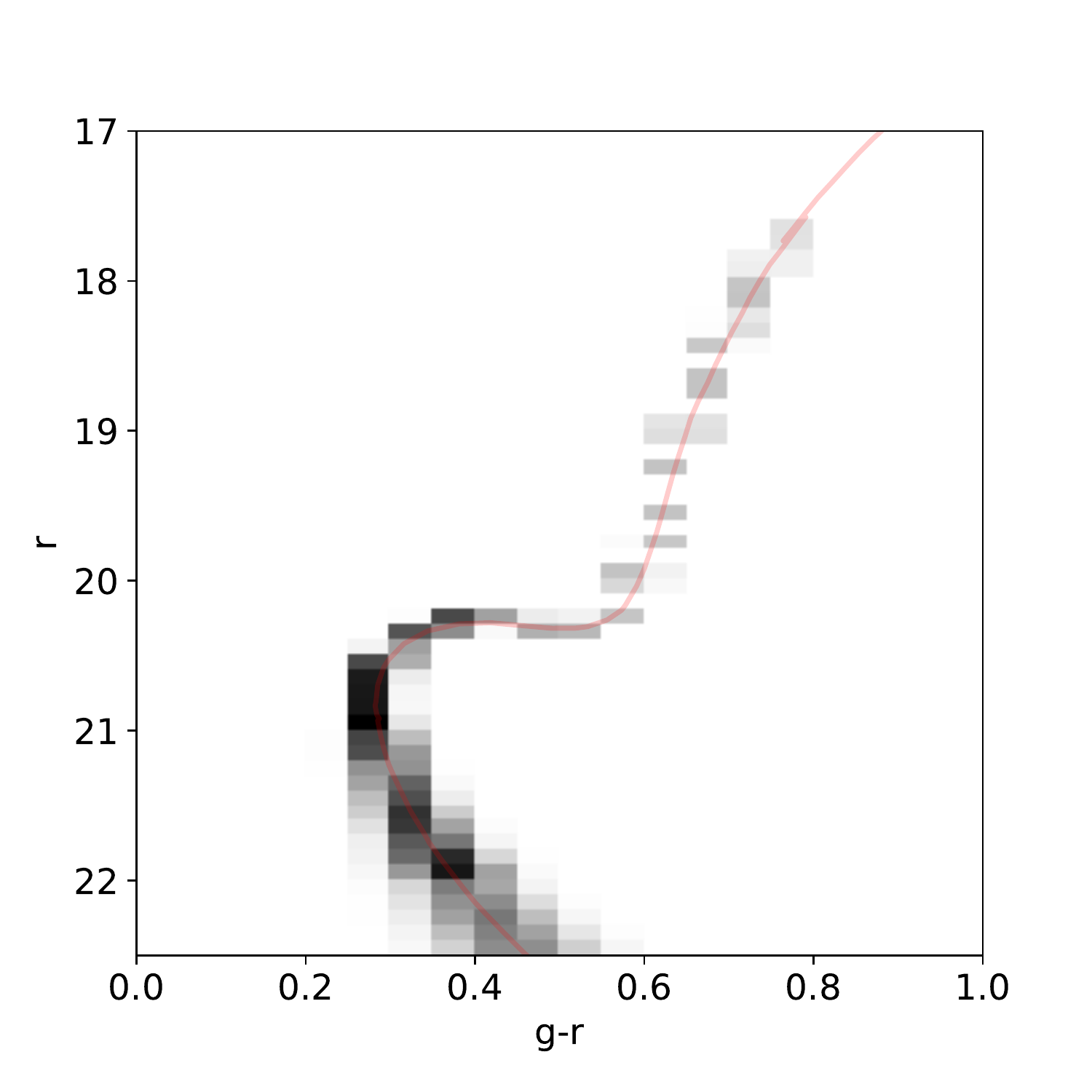}
\center
	\caption{The synthetic CMD filter for Whiting 1. It was constructed under the  
	the best-fit isochrone from Section \ref{isochrone} (red line in the plot), by considering the luminosity function of Whiting 1 and the typical observation errors. The bin size for the x-axis is 0.05 in 
	mag, and the bin size for the y-axis is 0.1 in mag. }
\label{fig_cmd}
\end{figure}

We apply the MF method to search extra-tidal features around Whiting 1. 
The basic procedures are described in \citet{2002AJ....124..349R}. Whiting 1 is embedded
in the Sgr trailing stream, so using an optimal CMD template is crucial for this work.
To build the filter, we have chosen to use synthetic CMD in $(g-r,r)$, which can utmost
avoid the contamination by field objects. We generate the synthetic CMD by using the best-fit isochrone
obtained in Section \ref{isochrone}. We convolve the isochrone with the luminosity
function of the cluster and apply the actual photometric uncertainties to the isochrone. 
No completeness correction is applied for the filter because we only use partial CMD ($r<$22.5 mag) as the MF template. The final CDM filter 
is shown in Figure \ref{fig_cmd}. 

\begin{figure*}
\center
	\includegraphics[width=1\textwidth]{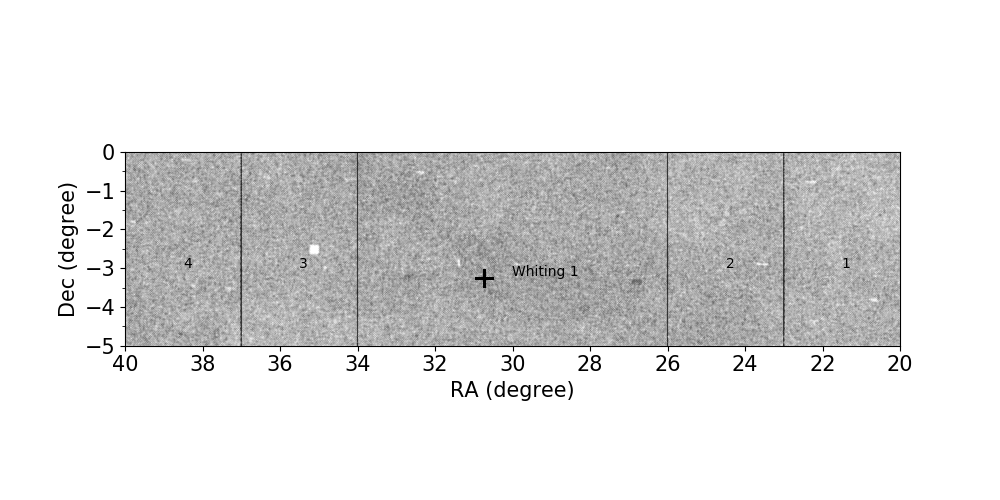}
\center
	\caption{Figure demonstrates how the Galactic background fields are selected. Fields 1-4 denote the star distribution of the background, defined at the boundaries: 20$^\circ$$<$RA$<$23$^\circ$, 23$^\circ$$<$RA$<$26$^\circ$, 34$^\circ$$<$RA$<$37$^\circ$, 37$^\circ$$<$RA$<$40$^\circ$, and
	-5$^\circ$$<$Dec$<$0$^\circ$. 
	The small blanks in field 1,2,3 are saturated or crowded areas with no public data. Plus symbol denotes the center of Whiting 1. The very faint diagonal feature around Whiting 1 is the Sgr trailing stream at where the cluster is embedded. }
\label{fig_background}
\end{figure*}

The Galactic background CMD is necessary for the MF method. An adjacent area of the sky is 
applied, excluding a window around Whiting 1. The four-region background is as shown in 
Figure \ref{fig_background}, with 
20$^\circ$$<$RA$<$23$^\circ$, -5$^\circ$$<$Dec$<$0$^\circ$, and 
23$^\circ$$<$RA$<$26$^\circ$, -5$^\circ$$<$Dec$<$0$^\circ$, and
34$^\circ$$<$RA$<$37$^\circ$, -5$^\circ$$<$Dec$<$0$^\circ$, and 
37$^\circ$$<$RA$<$40$^\circ$, -5$^\circ$$<$Dec$<$0$^\circ$. 
The average of the $(g-r,r)$ Hess diagrams is defined as the mean 
background density. 

The MF output, $\alpha$, which is defined as the densities of the stars that passed the MF
selection, is considered as the final signal of the substructure, as presented in Figure
\ref{fig_structure}. 

\section{Results}
\label{result}
\subsection{Extra-tidal features around Whiting 1}\label{features}
\begin{figure*}
\center
\includegraphics[width=0.42\textwidth]{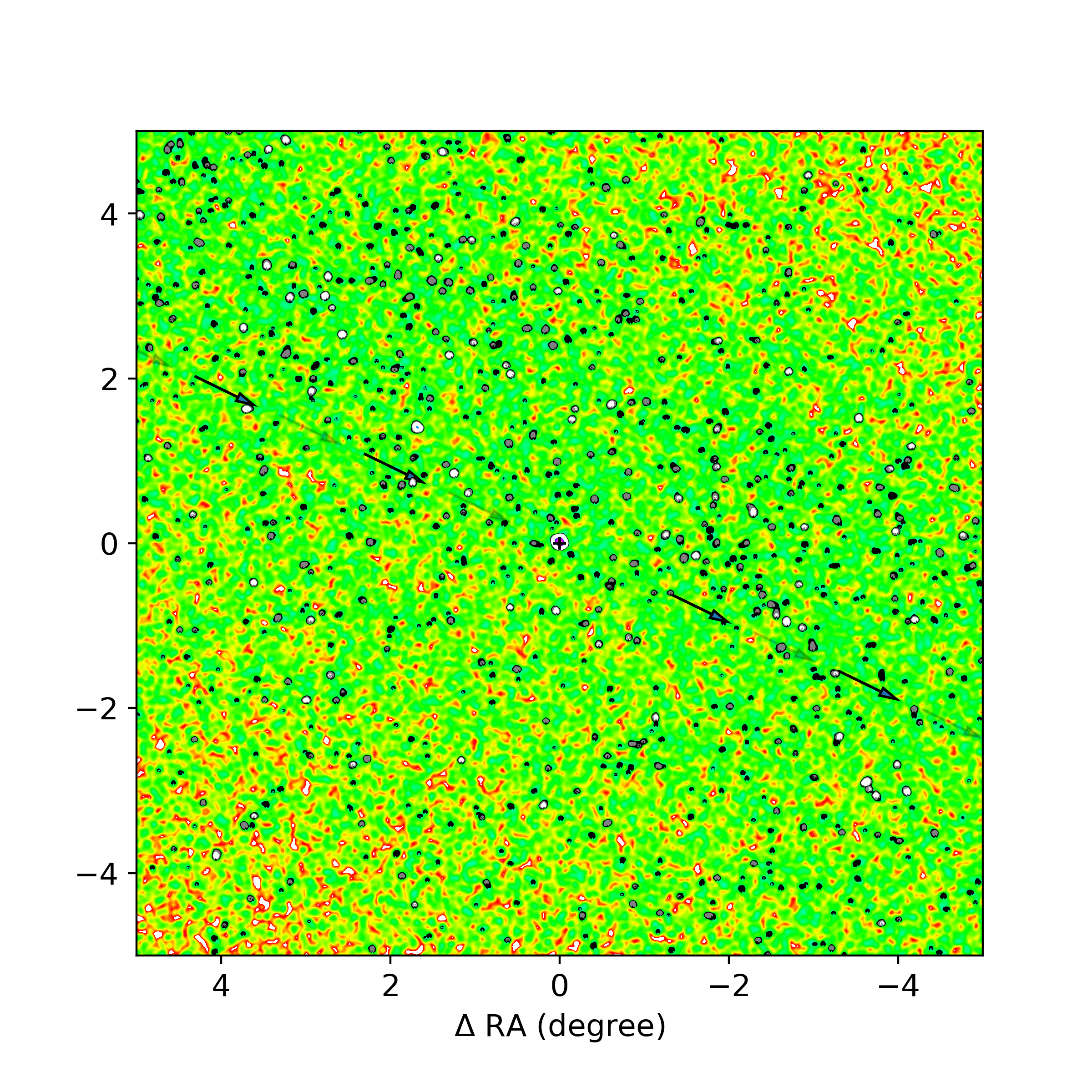}
\includegraphics[width=0.55\textwidth]{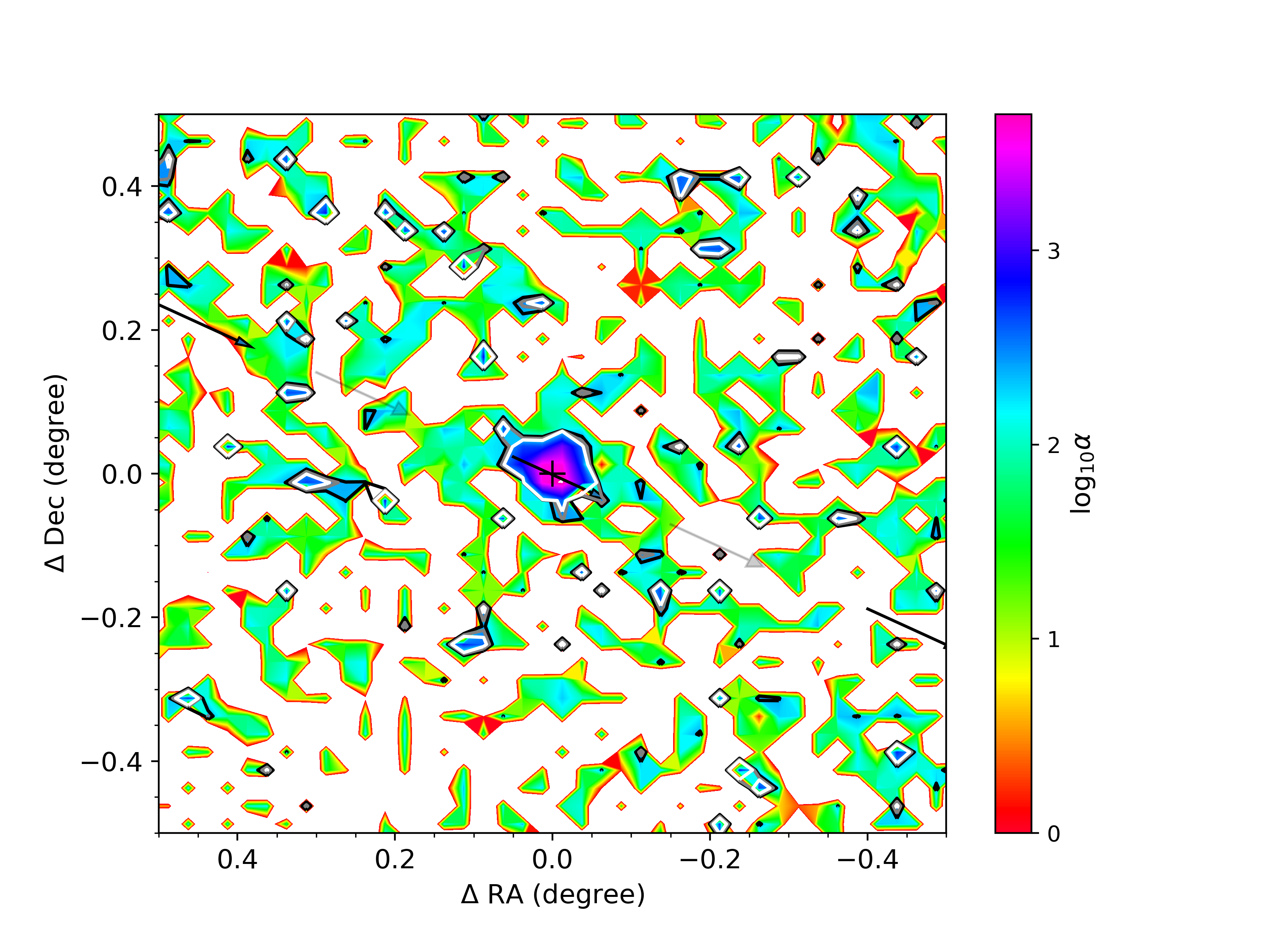}

\center
	\caption{The Logarithm of the density (log$_{10}\alpha$) of stars that 
	passed the matched filters $(g-r,r)$. The left panel is a smoothed density map in a field size of
	$10.0^{\circ}\times10.0^{\circ}$. The bin size of the density map is 0.025$^{\circ}$ and the smooth kernel is 1.0 pixel. The right panel is an unsmoothed density map, in a field
	size of $1.0^{\circ}\times1.0^{\circ}$. The bin size of the map is the same as that of the left panel. The center of each panel is Whiting 1, 
	as marked with a plus symbol. Color bar indicates the density of MF. Contour lines denote the 
	significance of the detected features, black for $3.0\sigma$, gray for $4.0\sigma$, 
	and white for $>5.0\sigma$. The gray arrow indicates the mean
	orbital direction of the nearby Sgr trailing stream according to \citet{2010ApJ...714..229L}. 
	The black arrow indicates the orbit direction of Whiting 1 as shown in Section \ref{orbit}.} 
\center
\label{fig_structure}
\end{figure*}

From the MF output, we obtain the distribution of $\alpha$ over the sky, 
which is directly relative to the numbers of stars that satisfy the CMD template
distribution. Figure \ref{fig_structure} shows the logarithm of the density
(log$_{10}\alpha$) centering in Whiting 1, with the left panel for a working
field of $10^{\circ}\times10^{\circ}$ and the right panel for a zoomed-in field of
$1.0^{\circ}\times1.0^{\circ}$. The bin size of the 
sky is 0.025$^\circ$. The left panel is smoothed by a Gaussian filter with
a standard deviation of 1 pixel, while for a detailed view of the real shape of the central feature, no smooth technique is applied to
the right panel. At the magnitude limit of $r$=22.5 mag, the survey completeness is 100\%,
so the density map is not affected by the spatial fluctuation. 
 
For each bin unit of Figure \ref{fig_structure}, we calculate its significance above 
the background. The background is at the average level of the field, excluding a window of 
$0.2^{\circ}\times0.2^{\circ}$ away from the Whiting 1 center. The background fluctuation 
is based on the standard deviation of the background region. The significance
levels are shown in contour lines in Figure \ref{fig_structure}, with black for 
$3.0\sigma$, gray for $4.0\sigma$, and white for $>5.0\sigma$. Since Whiting 1
is a faint and low-mass globular cluster, and it is embedded in the Sgr stream, 
very low S/N features are possibly not reliable. In this work, we are only 
care about the features with a significance higher than 3$\sigma$.

From the left panel of Figure \ref{fig_structure}, we can see many overdense
regions spreading over the whole working field. Among them, Whiting 1 is centered in the
field. The smoothed density map is easy for searching large-scale features extended from the
cluster center, e.g., something like the tidal tails of Palomar 5. However, we could not find
any large-scale tidal tails extending from
the cluster. Instead, there are many lumpy overdensities arbitrarily distributed all over the map.
We have also tried with different bin sizes for the density map, and they come to the same conclusion.
So we do not think there is a continuous, long tidal stream for Whiting 1. 

For a detailed view of the central feature, we zoom in the field in the right panel, but
without smoothing.
We see a round-shape feature with the size of a radius of 0.05$^{\circ}$-0.1$^{\circ}$ 
extended from the center of Whiting 1. Note that the tidal
radius of Whiting 1 is $r_{t}=$1.0$\arcmin$, but here we detected a much larger size for Whiting 1. Recalling the King-profile fit of the surface 
density profile of Whiting 1 \citep{2007A&A...466..181C}, our detection again
suggests there are extra-tidal stars around Whiting 1. From the deep
CFHT observations, \citet{2018ApJ...860...66M} suggested a tidal radius of 8.4$\arcmin$ for Whiting 1. This confirms a larger size of the cluster. Additionally, we see tiny tails extended from
either side of the cluster, which look like the trailing and leading tails of Whiting 1. Together with the central- round feature, the whole structure extends from ($\Delta$RA,$\Delta$Dec)=(0.05, 0.08) to (-0.07, -0.05).
By comparing to previous detections of Whiting 1, i.e., detections in
\citet{2017MNRAS.467L..91C} and \citet{2018MNRAS.476.4814S}, our result is 
consistent with the former detection but a little different to the later one. 
In fact, if we apply a smooth technique to our raw density map, it presents the same shape as that shown in \citet{2017MNRAS.467L..91C}. While to the detection of \citet{2018MNRAS.476.4814S}, the features of
\citet{2017MNRAS.467L..91C} and ours actually cover it: the detection of \citet{2018MNRAS.476.4814S} 
only shows a trailing tail on one side of the cluster, but our detection shows both the leading 
and trailing tails on opposite sides of the cluster. 

We also note that the tails are aligning along the orbital direction of the Sgr orbit as
well as that of the Whiting 1 orbit (gray and black arrows in the figure). The mean orbit of the Sgr
trailing arm is obtained according to \citet{2010ApJ...714..229L} model predictions within the range
$0^\circ<$RA$<90^\circ$. We overplot the Sgr orbit on Whiting 1 by doing a parallel shift. The Whiting 1 orbit is obtained from Section \ref{orbit}. From the figure, the
consistence between the tails and the orbital directions does not seem to be a coincidence, but
is likely to tell a picture of how Whiting 1 was formed: Whiting 1 was originally formed in the Sgr dSph,
and it fell into the Milky Way halo in groups, i.e., with dwarf galaxy. By the accretion of the Milky Way, it gradually grew into the current shape with leading and trailing tails on opposite sides of the cluster. 

\begin{figure*}
\center
\includegraphics[width=1\textwidth]{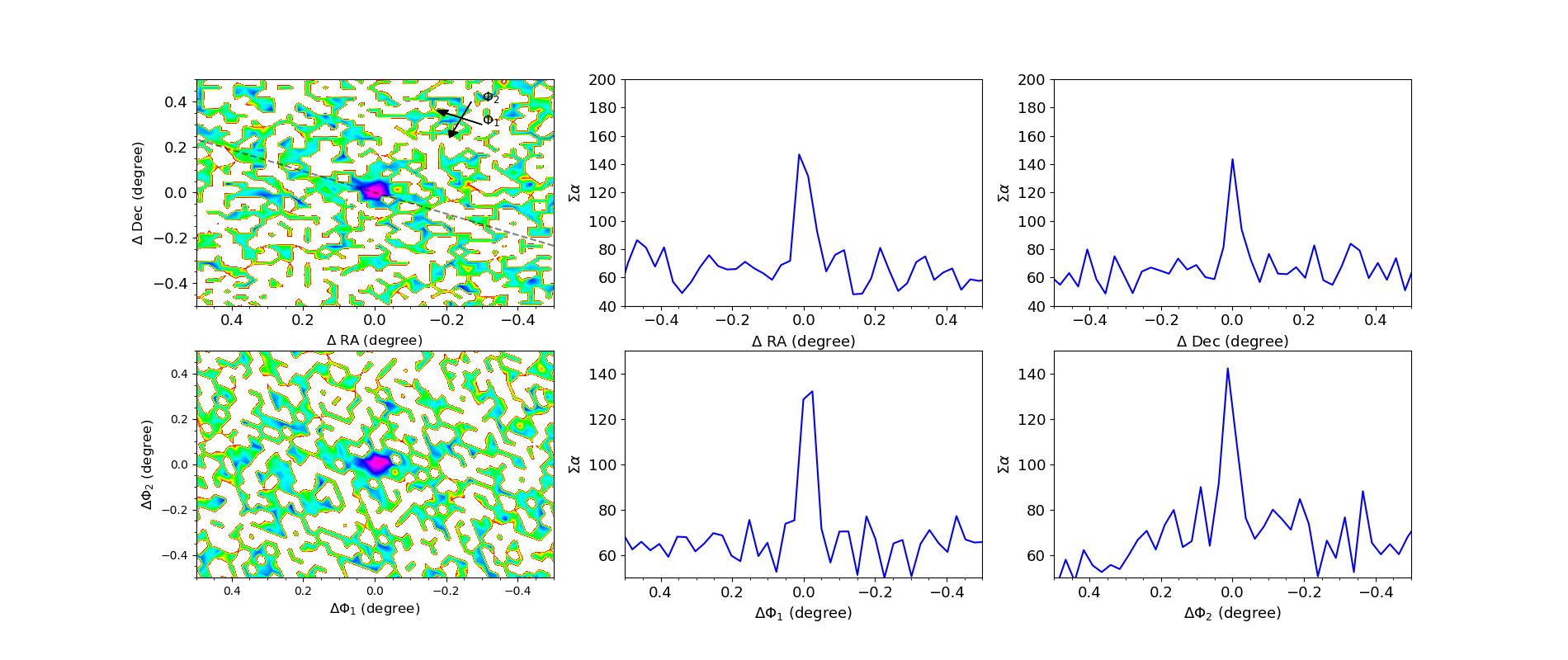}
\center
	\caption{Top panels: the density distribution of Whiting 1-like stars as a function of 
	$\Delta$RA and $\Delta$Dec. The coordinate reference for $\Delta$RA and $\Delta$Dec is
	the center of Whiting 1. The density in the middle and right panels is the integrated 
	intense of $\alpha$ along the x-axis. Bottom panels: the same as top panels but as a 
	function of $\Delta\Phi_{1}$ and $\Delta\Phi_{2}$. The coordinate reference for 
	$\Delta\Phi_{1}$ and $\Delta\Phi_{2}$ is the center of Whiting 1. 
	The transformation from (RA,Dec) to ($\Phi_{1}$,$\Phi_{2}$) is indicated in the top 
	left panel. The transformation follows the orbit of Sgr dSph defined in \citet{2010ApJ...714..229L}. }
\center
\label{fig_alpha_dist}
\end{figure*}

To verify the reality of whole central feature, which is from ($\Delta$RA,$\Delta$Dec)=(0.05, 0.08) to (-0.07, -0.05), we check the density distribution of stars
that passed the MF (defined as $\alpha$). The top panels of Figure~\ref{fig_alpha_dist} show the density distribution of Whiting 1-like stars as a function of 
$\Delta$RA and $\Delta$Dec. It is clearly seen that there is a density enhancement around the center of the
cluster. For a significance $>3\sigma$, the width
of this overdensity is about 0.1$^{\circ}$-0.2$^{\circ}$. This confirms the reality of 
the central feature. If we excluding the Whiting 1 core stars in the very center, we can still find an overdensity from -0.1$^{\circ}$ to 0.1$^{\circ}$ along $\Delta$RA or $\Delta$Dec,
which means the $>3\sigma$ overdensity is not only contributed by the core stars of Whiting 1 but also extra-tidal stars of Whiting 1.
As the whole central feature aligns along the orbital direction of the Sgr dSph, we did a rotation to
make the $x$-axis of the top left panel along the orbital direction of the Sgr dSph, as shown in
the bottom left panel of Figure~\ref{fig_alpha_dist}. Then we analyze the density distribution of
Whiting 1-like stars along the orbital direction of $\Delta\Phi_{1}$ and $\Delta\Phi_{2}$. As expected, we see 
similar density enhancements along $\Delta\Phi_{1}$ and $\Delta\Phi_{2}$. The width of the overdensity is about
0.1$^{\circ}$-0.2$^{\circ}$ for a significance $>3\sigma$.
The density distributions of Whiting 1-like stars indicate the reality of the central overdensity.

\subsection{The characteristics of the extra-tidal features}
The whole view of Whiting 1 structure draws our attention for 
two reasons:(1) Its size is larger than the 
cluster core. (2) Its shape is not exactly 
round; it has tails on opposite directions of the cluster with an elongation aligning
along the orbital direction of Whiting 1 as well as the orbital direction of the Sgr trailing stream. We analyze the characteristics of this feature by studying stars inside
and outside the structure region.

\begin{figure*}
\center
\includegraphics[width=0.99\textwidth]{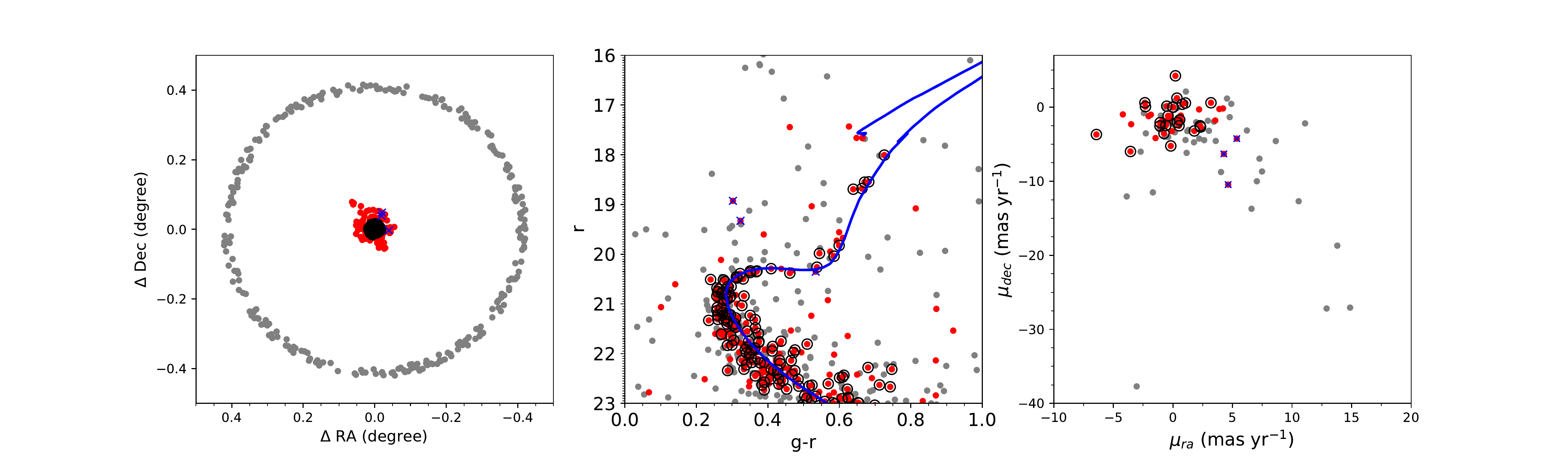}
\center
	\caption{Left panel: spatial distribution of stars inside and outside the Whiting 1
	structure. The coordinate reference is the center of Whiting 1. Red dots denote stars inside the Whiting 1 structure in Figure \ref{fig_structure}, gray dots denote stars outside the Whiting 1 structure, which are 0.4$^{\circ}$ away from the Whiting 1 center. The black circles are Whiting 1 core stars.  The blue solid line is the Whiting 1 CMD isochrone from Figure \ref{fig_isochrone}. Middle panel: the CMD distribution of the stars inside and outside the structure. Right panel: the proper-motion distribution of the stars 
	inside and outside the structure. The proper-motion data are from Gaia EDR3. Only 16\% of the inside or outside- stars have proper-motion data, so the distribution is incomplete. The blue crosses denote inside-stars that have proper motion largely away from the
	proper-motion clump.}
\center
\label{fig_s0_inf}
\end{figure*}

The left panel of Figure \ref{fig_s0_inf} shows how we select the study sample. 
The red dots in Figure \ref{fig_s0_inf} are stars inside the whole feature without
any rejections. The gray dots in the figure denote the comparing stars, which are
0.4$^{\circ}$ away from the Whiting 1 center. We constrain the number of the comparing stars 
in an annulus with the same number-counts as that of the inside-stars.  
The black circles in the figure denote the Whiting 1 core stars, which are inside a tidal radius of 1.0$\arcmin$.
The middle panel shows the CMD distribution for stars inside and 
outside the structure. As seen, most (92\%) of the inside-stars follow the CMD distribution
of the core stars, while only 46\% of the outside-stars follow, with the rest of them 
randomly distributed in the CMD. For the 46\% population, most of them are faint stars. Considering the poor accuracy at the faint end, 
we suggest that the inside-stars are highly Whiting 1 members while the outside-stars are probably
from the Sgr trailing stream. 

We also tried to study the kinematics of the inside- and outside-stars, however, only 16\% 
of them have velocity data, i.e., proper-motion from Gaia EDR3. In the right panel of Figure
\ref{fig_s0_inf}, we exam the proper-motion distribution of the core stars, 
and found the core stars roughly follow a distribution with a median of ($\mu_{\alpha}$,$\mu_{\delta}$)=
(-0.1,-2.3) mas/year, and a dispersion of 4.0 mas/year. This distribution
is close to the proper motion provided by \citet{2019MNRAS.482.5138B} and 
\citet{2021MNRAS.505.5957B}, results of which are based on Gaia
DR2 and EDR3. For most of the inside-stars (except three stars marked with blue crosses), they follow
the proper-motion distribution of the core stars. While for the outside-stars, a large fraction of them are away from the proper motion clump.

As we know, the Gaia EDR3 proper motion is only accurate down to G=21 mag, so for most of
the stars inside or outside the structure, especially the faint stars, their proper motions are not
available or have large errors. Owing to this, the right panel cannot tell us the complete
proper-motion distribution. We show the proper motion here just for a reference. We need more deep velocity data
to study their kinematics. Hopefully, in the future, we can get more velocity data with much deeper
spectroscopic surveys, such as the China Space Station Telescope (CSST) and DESI.

\subsection{Consider possible contamination to the detection}
\begin{figure*}
\center
\includegraphics[width=.95\textwidth]{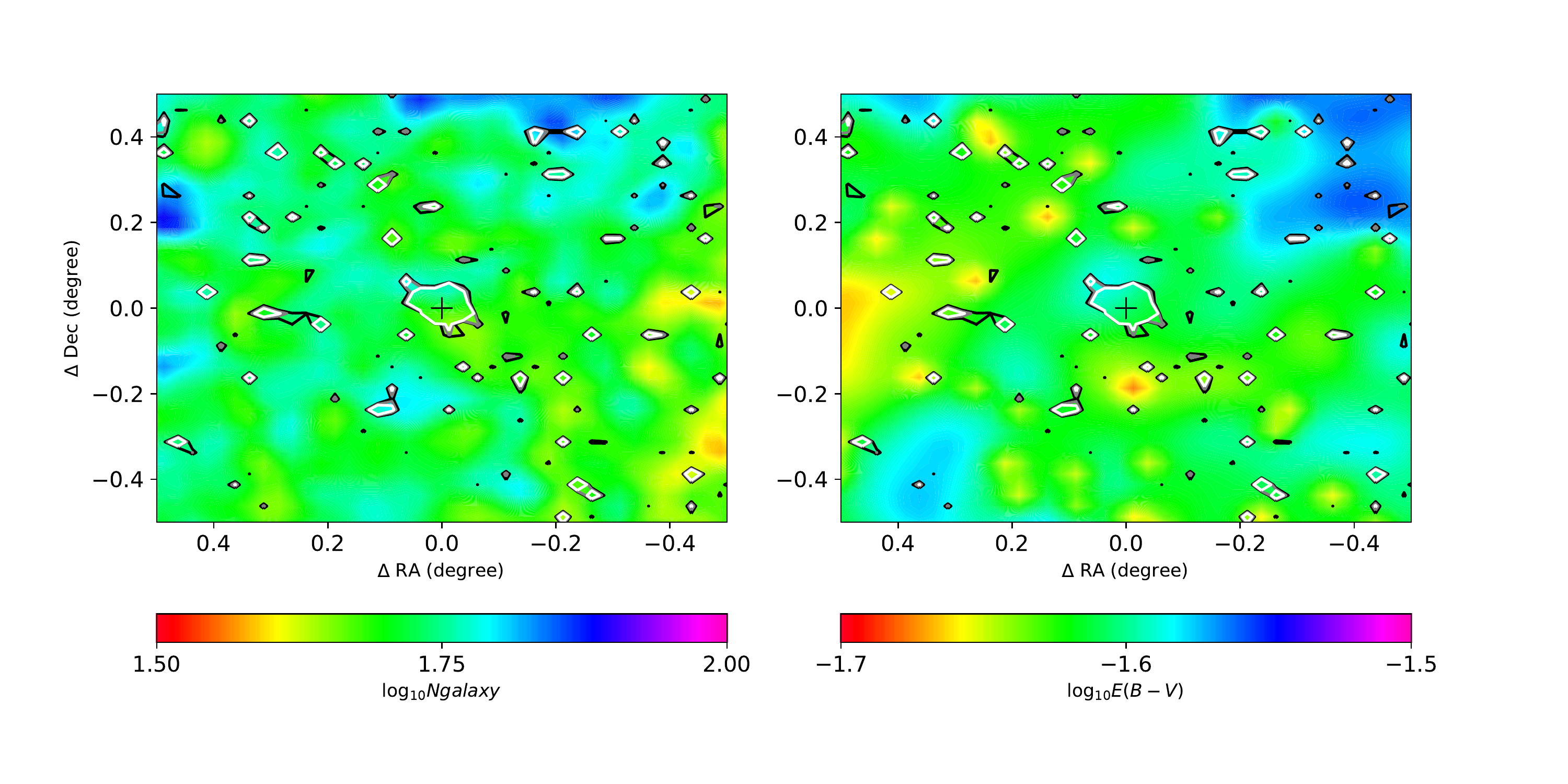}
\includegraphics[width=.95\textwidth]{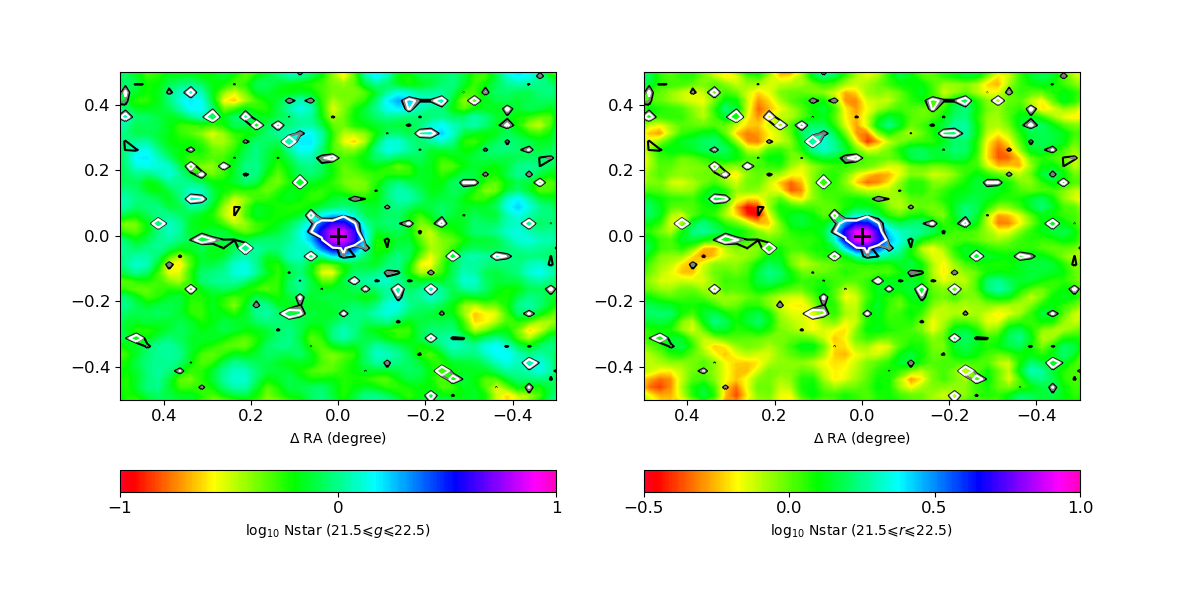}
\center
	\caption{Top left panel: number-density distribution of the galaxies around Whiting 1.
	Top right panel: the E(B-V) extinction distribution around Whiting 1. Bottom left and right panels: the number-density 
	distributions for stars in a magnitude interval of 21.5-22.5 mag for $g$ and $r$ bands, respectively. To guide eyes, the MF detections with
	significance $>3\sigma$ are overplotted with contour lines, with black for 3$\sigma$, gray for 4$\sigma$, white for $>5\sigma$.  Pixel scale of the four
	panels is the same as in Figure \ref{fig_structure}. }
\center
\label{fig_ebv}
\end{figure*}

As we are using the deeper sample from DECaLS DR9, there are many galaxies in the 
relative sky coverage. To check that the background galaxies are not responsible 
for our detected substructures, we present their density distribution by using 
the stellar/galaxy classification provided in DECaLS DR9 catalog, i.e., sources 
with COMP, DEV,REX, and EXP magnitude types are considered as galaxies. The 
comparison between the galaxy density distribution and our substructions is presented in the left panel of Figure \ref{fig_ebv}. As seen, no 
correlation is found between them. Another possible mimic of the 
substructures is the interstellar extinction. We show the dust map in the right panel of 
Figure \ref{fig_ebv}. We find that there is no strong
correlation between the substructures and the dust map. 

In addition, to check the contamination of faint stars, we examine their number-
density distribution. We compare the density map of faint stars, i.e., stars with $g$
or $r$ magnitude between 21.5 and 22.5 mag, to our MF detection, and we show the results in the
bottom panels of Figure \ref{fig_ebv}. This comparison tells us that the faint stars
do contribute to the central feature. 
However, when we go to a bright-magnitude interval, e.g., $r$ between  
19 and 21.5 mag, or other brighter-magnitude intervals, 
the central structure is still there and shapes the same. This
means that the shape and size of the central feature are real, and they are not influenced by the faint
stars.

\section{Summary}\label{summary}
We apply the MF method on the deep survey DECaLS data to search for
extra-tidal features around the Whiting 1 cluster. Considering the completeness of the
data, we constrain the working sample with $r<=$22.5 mag. Without any smooth technique, 
we detected a round-shape feature centering in Whiting 1. This central feature has a size in a radius of 0.05$^{\circ}$-0.1$^{\circ}$, which gives an area much larger than the Whiting 1 core. Our detection also
shows tiny tails in the opposite direction of the central feature, which looks like the leading and
trailing tails of Whiting 1. The tails
align well along the orbital motion of Whiting 1 and the mean orbital direction of the Sgr
trailing stream. To investigate the origin of the whole feature,  
we checked stars inside the feature with their CMD and proper motions. Most of the stars 
follow the distributions of the Whiting 1 core stars, which infers that the whole feature
possibly belongs to Whiting 1. 

The whole view of detected feature is consistent with the detection of
\citet{2017MNRAS.467L..91C} and covers the detection of \citet{2018MNRAS.476.4814S}.
Even though we did not find new features compared to the previous study, our detection is
the raw density distribution without any smoothing; we can trust on its shape and size. We infer that
the detected feature is possibly the true face of Whiting 1 under a depth of $r<22.5$ mag.
The morphology of the whole feature seems to suggest that Whiting 1 was 
initially born in the Sgr dSph, and gradually grew into the current shape by the accretion 
of the Milky Way. The intriguing origin of Whiting 1 makes it to be a perfect example to
study the formation of the Milky Way clusters.

\begin{acknowledgments}
The authors are very grateful to Prof.Vasily Belokurov and Prof. Carl Grillmair for constructive 
suggestions and helpful discussion. J.N. acknowledges the supports 
by the National Key R\&D Program of China (grants No. 2021YFA1600400,2021YFA1600401,2019YFA0405501), by the science
research grants from the China Manned Space Project (grants NO.
CMS-CSST-2021-B03,CMS-CSST-2021-A10), by the Chinese National 
Natural Science Foundation (grants No. 12073035,12173051,11873053,11633002,12120101003).
H.T. acknowledges support from the National Key R\&D Program of China under grant No. 2019YFA0405500. Jing Li acknowledges the support from the National Natural Science Foundation under
grant No.11703019. B.T. acknowledges the support from the National Natural Science Foundation 
of China under grant No.U1931102  and the support from the hundred-talent project of Sun Yat-sen
University. H.T. acknowledges the support from the National Natural Science Foundation of 
China under grant No.11873034.

The DESI imaging Legacy Surveys consist of three individual and complementary projects: the 
Dark Energy Camera Legacy Survey (DECaLS; NOAO Proposal ID $\#$ 2014B-0404; PIs: David Schlegel 
and Arjun Dey), the Beijing-Arizona Sky Survey (BASS; NOAO Proposal ID $\#$ 2015A-0801; PIs: Zhou Xu 
and Xiaohui Fan), and the Mayall z-band Legacy Survey (MzLS; NOAO Proposal ID $\#$ 2016A-0453; 
PI: Arjun Dey). DECaLS, BASS, and MzLS together include data obtained, respectively, at the 
Blanco telescope, Cerro Tololo Inter-American Observatory, National Optical Astronomy 
Observatory (NOAO); the Bok telescope, Steward Observatory, University of Arizona; and the 
Mayall telescope, Kitt Peak National Observatory, NOAO. The Legacy Surveys project is honored 
to be permitted to conduct astronomical research on Iolkam Du'ag (Kitt Peak), a mountain with 
particular significance to the Tohono O$_{'}$odham Nation.

This work presents results from the European Space Agency (ESA) space mission Gaia. Gaia data are being processed by the Gaia Data Processing and Analysis Consortium (DPAC). Funding for the DPAC is provided by national institutions, in particular the institutions participating in the Gaia MultiLateral Agreement (MLA). The Gaia mission website is https://www.cosmos.esa.int/gaia. The Gaia archive website is https://archives.esac.esa.int/gaia.
\end{acknowledgments}

\end{document}